\documentclass[amssymb,nobibnotes,superscriptaddress,longbibliography,notitlepage,twocolumn]{revtex4-1}
\usepackage{enumerate,amsmath}
\usepackage[dvipdfmx]{graphicx}
\usepackage[dvipdfmx]{color} 

\newtheorem{theo}{Theorem}

\newtheorem{lemm}{Lemma}

\newcommand{\mU}{\mathcal{U}}

\newcommand{\mN}{\mathcal{N}}
\newcommand{\mI}{\mathcal{I}}
\newcommand{\mE}{\mathcal{E}}

\begin{document}

\title{Noise Threshold of Quantum Supremacy}%

\author{Keisuke Fujii}
%\email{fujii.keisuke.2s@kyoto-u.ac.jp}
\address{Photon Science Center, Graduate School of Engineering,
The University of Tokyo, 2-11-16 Yayoi, Bunkyo-ku, Tokyo 113-8656, Japan}
\address{JST, PRESTO, 4-1-8 Honcho, Kawaguchi, Saitama, 332-0012, Japan}

\date{\today}
\begin{abstract}
Demonstrating quantum supremacy, a complexity-guaranteed quantum advantage 
against over the best classical algorithms by using less universal quantum devices,
is an important near-term milestone for quantum information processing.
Here we develop a threshold theorem for quantum supremacy with
noisy quantum circuits in the pre-threshold region,
where quantum error correction does not work directly.
We show that, even in such a region, 
we can virtually simulate quantum error correction 
by postselection.
This allows us to show that the output sampled from
the noisy quantum circuits (without postselection)
cannot be simulated efficiently by classical computers 
based on a stable complexity theoretical conjecture, i.e., non-collapse of the polynomial hierarchy.
By applying this to fault-tolerant quantum computation with the surface codes,
we obtain the threshold value $2.84\%$ for quantum supremacy,
which is much higher than the standard threshold $0.75\%$ for universal fault-tolerant quantum computation
with the same circuit-level noise model.
Moreover, contrast to the standard noise threshold,
the origin of quantum supremacy in noisy quantum circuits
is quite clear; the threshold is determined purely by 
the threshold of magic state distillation,
which is essential to gain a quantum advantage.
\end{abstract}

%\pacs{}
\maketitle
\section{Introduction}
One of the most important goals of quantum information processing
is to demonstrate quantum speedup over the best classical algorithms,
namely quantum supremacy~\cite{Yao03,Preskill12,Aaronson11}
to disproof the extended Church-Turing thesis, saying that
any efficient computation by a realistic physical device can be efficiently simulated by a probabilistic Turing machine. 
For example, if we have an ideal universal quantum computer,
Shor's factorization algorithm~\cite{ShorFactoring} allows us to 
demonstrate a super-polynomial quantum speedup over 
the best known classical algorithms.
Moreover, the threshold theorem for fault-tolerant quantum computation guarantees
that an ideal universal quantum computer
can be constructed from realistic quantum physical devices 
being subject to physically realistic imperfections~\cite{KitaevThreshold,PreskillThreshold,KnillThreshold,AharonovBen-OrThreshold}.
Therefore, massive efforts have been paid 
for experimental realizations of fault-tolerant quantum error correction~\cite{Martinis14,Martinis15,Martinis15b,IBM14,IBM15,IBM16,QuTech15}.

Recently intermediate models of quantum computation
are attracting much attention to show quantum supremacy 
in experimentally feasible settings.
They are not rich enough to perform 
universal quantum computation,
but still provide nontrivial outputs,
which could not be simulated efficiently by classical computers.
BosonSampling with free bosons~\cite{Aaronson11}, 
instantaneous quantum polynomial-time computation (IQP) with commuting quantum circuits~\cite{IQP0,IQP1,IQP2,IQP3,IQP4,IQP5} (see also depth-four circuits~\cite{Depth4}), 
highly mixed deterministic quantum computation with one-clean qubit (DQC1)~\cite{KnillLaflamme,MorimaeFF,KKMNTT}
are examples of those.
All of them are experimentally well motivated 
as linear optical quantum computations~\cite{KLM}, quench dynamics with Ising interactions~\cite{Duan16,Farhi16},
and NMR ensemble quantum computation~\cite{KnillLaflamme}.
Specifically, it has been shown that 
if the output of these intermediate models are 
sampled efficiently by a classical computer, the polynomial hierarchy (PH),
a generalization of NP (nondeterministic polynomial-time computation) to oracle machines,
collapses to the third (or second~\cite{KKMNTT}) level~\cite{Aaronson11,IQP1}.
The collapse of the PH is thought to be highly implausible (for example, P=NP implies that a complete collapse of the PH),
and hence classical simulation of the intermediate models is also thought to be hard.
Based on this understanding, 
several BosonSampling experiments have been performed already~\cite{BSex1,BSex2,BSex3,BSex4,BSex5,BSex6,BSex7,BSex8}.

Here we consider quantum supremacy of noisy quantum circuits
in the pre-threshold region, 
where the noise strength is much higher than 
the standard threshold of universal fault-tolerant quantum computation.
Hence, we cannot employ quantum error correction directly.
Then we ask whether or not there is still surviving quantum supremacy
in such noisy quantum circuits.
The motivation of this question is threefold. 
(i) There have been
several noise thresholds above which such noisy quantum circuits
are classically simulatable exactly.
However, they are assuming ideal stabilizer operations~\cite{Buhrman06,Howard09},
or there still be a large gap to the threshold of
universal fault-tolerant quantum computation~\cite{Plenio05,Plenio10,Barrett09}.
The characterization of the intermediate pre-threshold region,
where the standard quantum error correction does not work,
has been fully open for a long time. 
(ii) The hardness proofs of the existing intermediate models
require the sampling with constant multiplicative errors 
or constant additive errors with $l_1$-norm~\cite{Aaronson11,IQP1,IQP2,IQP4,IQP5}.
These notions of approximation
are quite sensitive to noise.
If noisy quantum circuits of a constant noise strength 
are employed, these criteria cannot be achieved directly.
Thus, whether or not such noisy quantum circuits themselves can 
exhibit quantum supremacy has been an important open problem (see also Ref.~\cite{IQP3}).
(iii) Nowadays, it is becoming possible to operate scalable quantum devices
such as superconducting qubits
around the standard threshold of universal fault-tolerant quantum computation~\cite{Martinis14,Martinis15,Martinis15b,IBM14,IBM15,IBM16,QuTech15}.
Therefore, a complexity-guaranteed criterion of quantum supremacy for such noisy quantum devices in
the pre-threshold region
is highly demanded in the experiments~\cite{Boixo16}.
%to falsify
%the extended Church-Turing thesis 
%with a realistically noisy quantum devices.

To address these issues, we derive 
a threshold theorem for quantum supremacy with noisy quantum circuits:
if the noise strength is lower than a certain threshold value,
the output of such noisy quantum circuits cannot be simulated 
efficiently by classical computers unless the PH hierarchy collapses to the third level.
Contrast to the existing intermediate models
~\cite{Aaronson11,IQP0,IQP1,IQP2,IQP3,IQP4,IQP5,MorimaeFF,KKMNTT},
in this work we employ quantum circuits generated from a universal set of gates,
but being subject to rather strong noise,
which makes the system less universal.
To show hardness of classical simulation,
we employ the postselection argument~\cite{postBQP,IQP1}.
That is, we show that
noisy quantum circuits 
can solve a postBQP-complete, or equivalently 
PP-complete problem under postselection.
To this end, we first show that
simulation of
an arbitrary two-qubit output of universal quantum computation
with an exponentially small additive error
is enough to obtain the hardness result by postselection.
Then, 
we show that noisy quantum circuits can achieve it 
by virtue of postselection,
where any outcomes of syndrome measurements suggesting 
existence of errors are discarded by postselection.
This allows us to simulate universal quantum computation
with an exponentially small additive error
even in the pre-threshold region,
where the standard quantum error correction is not available.
As a technical point of view,
the standard threshold theorem for universal fault-tolerant quantum computation 
cannot be employed in the above argument.
This is because, in the postselection argument,
we have to treat a conditional probability distribution
conditioned on an exponentially rare postselection event.
Therefore we derive a postselected version of the threshold theorem.

The important implications of the postselected version of the threshold theorem
are as follows.
First, while the raw outputs of noisy quantum circuits
cannot satisfy the criteria for quantum supremacy directly,
the logical output after an appropriate classical processing 
can exhibit quantum supremacy by virtue of quantum error correction,
which is virtually simulated by using postselection.
Second, the threshold value for quantum supremacy is 
much higher than that of universal fault-tolerant quantum computation.
This is because, we can discard any erroneous events suggested by
the non-zero error syndromes. Therefore the threshold is 
given not by the error correction property of quantum error correction codes 
but by the error detection property. 
%That is, a quantum error correction code 
%of distance $d$ can correct only up to weight-$\lfloor (d-1)/2\rfloor$ errors 
%but can even detect weight-$(d-1)$ errors.
Third, by virtue of the above effect,
the origin of quantum supremacy in noisy quantum circuits is
quite clear; it is determined by distillability of the magic state~\cite{MSD1,MSD2}.
More precisely, 
in the standard construction of fault-tolerant quantum computation,
error correction for Clifford gates
limits the threshold,
while the threshold of magic state distillation,
where error detection is employed, is 
much higher~\cite{RaussendorfAnn,RaussendorfNJP,RaussendorfPRL,Fowler09}.
In the case of quantum supremacy,
we can employ error detection for both Clifford gates and magic state distillation,
and hence distillability of the magic state 
determines the threshold of quantum supremacy.
%Clifford gates and non-Clifford gates with magic state distillation
%are protected in different ways.
%Specifically, magic state distillation~\cite{MSD1,MSD2} for non-Clifford gates operates 
%via error detection, since we can adaptively chose 
%successfully distilled magic states.
%On the other hand, Clifford gates are protected by 
%error correction.
%The threshold for error correction is smaller than
%that for error detection in general,
%and hence the standard threshold for universal fault-tolerant
%quantum computation is limited by the threshold for protecting Clifford gates~\cite{RaussendorfAnn,RaussendorfNJP,RaussendorfPRL,Fowler09}.
%However, in the case of the postselected threshold theorem,
%we can employ error detection both Clifford gates and magic state distillation.
%As a result, the threshold for distillability of magic states 
%determines the threshold for quantum supremacy.
This is quite reasonable, since magic state distillation 
for non-Clifford gates is essential for make quantum computation 
classically intractable~\cite{Howard14,Buhrman06,Howard09,Plenio10,Howard14}.
Finally, we calculate the threshold value 
of fault-tolerant quantum computation using the surface code
on the two-dimensional array of qubits~\cite{RaussendorfPRL,Fowler09}
to obtain a practically meaningful threshold of quantum supremacy.
While we cannot simulate postselected events numerically 
and hence employ an analytical treatment, which 
underestimates the threshold,
the resultant threshold value $2.84\%$ is rather high 
compared to the standard threshold $0.75\%$~\cite{RaussendorfAnn,RaussendorfNJP}
under the same circuit-based depolarizing noise model.
This level of noise is within reach of current state-of-the-art
experiments in scalable superconducting qubit systems~\cite{Martinis14,Martinis15,Martinis15b,IBM14,IBM15,IBM16,QuTech15},
and hence it would be possible to observe complexity-guaranteed quantum supremacy
with noisy quantum circuits in the near future.

The rest of the paper is organized as follows.
In Sec.~\ref{sec:01},
we briefly review the standard threshold theorem
of universal fault-tolerant quantum computation.
In Sec.~\ref{sec:02},
we construct the postselected threshold theorem
of quantum supremacy with noisy quantum circuits.
In Sec.~\ref{sec:03},
we perform a case study for concatenated quantum computation.
In Sec.~\ref{sec:04},
we apply the postselected threshold theorem
to topological fault-tolerant quantum computation 
with the surface codes to derive the 
practical threshold value of quantum supremacy
with the circuit-level noise.
Section~\ref{sec:05}
is devoted to conclusion and discussion.

\section{Standard threshold theorem}
\label{sec:01}
Let us briefly review the standard threshold theorem of 
fault-tolerant quantum computation~\cite{AharonovBen-OrThreshold,Aliferis}. 
The ideal unitary gates in a fault-tolerant universal quantum computation 
is denoted by 
\begin{eqnarray}
\mathcal{U} = \prod _k \mU _k, 
\end{eqnarray}
where $\mU _k$ is the $k$th unitary gate and 
is chosen from a universal set of gates.
Quantum error correction employs 
both projective measurements and adaptive operations
based on the measurement outcomes.
Here, for simplicity, all these operations 
including classical processing
are denoted by unitary gates coherently
and included in $\mU $.
Then, the probability distribution of 
the final output of quantum computation 
is denoted in terms of a projective operator $P_x$ ($x \in \{ 0,1\}$)
of the final readout
as 
\begin{eqnarray}
p(x) = {\rm Tr}[P_x \mU (\rho _{\rm ini})],
\end{eqnarray}
where $\rho _{\rm ini}$ is the initial state.
In order to take noise into account,
each ideal unitary gate $\mU_k$ 
is replace with a noisy one $\mN_k \mU_k$,
where $\mN_k$ is a completely-positive-trace-preserving (CPTP) map
representing the imperfection.
Here we assume the noise is local and Markovian.
We define a noise strength $\epsilon _k \equiv \| \mI - \mN_k \|_{\diamond}$ 
for each $\mN_k$,
where $\| \cdot \|_{\diamond}$ is the diamond norm 
for super-operators~\cite{diamondnorm}.
Then the noisy version of the unitary gates 
is given by
$\mU ^{\rm noisy} \equiv \prod _{k} \left(\mN_k \mU_k \right)$.
Note that imperfections on the initial state $\rho _{\rm ini}$
and the final measurement $P_x$ are also taken as 
imperfections on unitary gates at certain locations
(this is always possible since we can insert an identity gate after the state preparation and before the measurement).
Note also that 
the unitary gates corresponding to the classical processing 
are assumed to be noise-free, since they are introduced just to simplify the argument.
They are implicitly omitted in the following argument.
The probability distribution of the 
final output under the noise is given by 
\begin{eqnarray}
p^{\rm noisy}(x) = {\rm Tr}[ P_x \mU^{\rm noisy} (\rho_{\rm ini})].
\end{eqnarray}
To argue fault-tolerance,
we decompose the noise map $\mN _k$ 
into ideal and noisy parts as follows:
\begin{eqnarray}
\mN_k = (1-\epsilon_k)\mI + \mE_k
\end{eqnarray}
where we should note that the noisy part $\mE_k$ (with $\| \mE \|_{\diamond} \leq 2\epsilon_k$) is 
not always a CPTP map.
Then we expand $\mU^{\rm noisy}$ as
a summation over possible paths 
\begin{eqnarray}
\mU^{\rm noisy} &=& \prod _k \left\{ \left[ (1-\epsilon_k)\mI + \mE_k \right] \mU_k \right\}
\\
&=& \sum _{\{ \eta _k \}} \prod _{k} \left\{ \left[ (1-\epsilon _k) \mI \right]^{1-\eta _k} \mE_k^{\eta _k} 
\mU_k \right\},
\end{eqnarray}
where $\eta _k \in \{ 0,1\}$ and $\sum _{\{ \eta _k \}}$ 
is taken over all paths.
Now we decompose these paths into 
sparse and faulty set of paths
in such a way that  
the operator in the sparse set
never change the final probability distribution:
\begin{eqnarray}
p(x) &\propto& {\rm Tr}\left[ P_x 
\sum _{\{ \eta _k \}| {\rm sparse} } 
\prod _{k} \left\{ \left[ (1-\epsilon _k)\mI \right]^{1-\eta _k} \mE^{\eta _k} \mU_k \right\} \rho _{\rm ini} \right] 
\nonumber \\
&&\\
&\equiv& \alpha p(x).
\end{eqnarray}
The faulty set is defined as the complement of the sparse set.
The sparse and faulty operators (not a density operator)
are defined as follows:
\begin{eqnarray}
\rho _{\rm faulty} &\equiv& \sum _{\{ \eta _k \}| {\rm faulty} } 
\prod _{k} \left\{ \left[ (1-\epsilon _k)\mI \right]^{1-\eta _k} \mE^{\eta _k} \mU_k \right\} \rho _{\rm ini},
\\
\rho _{\rm sparse} &\equiv& \sum _{\{ \eta _k \}| {\rm sparse} } 
\prod _{k} \left\{ \left[ (1-\epsilon _k) \mI\right]^{1-\eta _k} \mE^{\eta _k} \mU_k\right\} \rho _{\rm ini}.
\end{eqnarray}
The error of the probability distribution of the final output 
is measured by $l_1$-norm,
\begin{eqnarray}
\| p(x)-p^{\rm noisy}(x) \|_{l_1} 
= \| (1- \alpha)p(x) + {\rm Tr}[P_x \rho _{\rm faulty} ]\|_{l_1}.
\nonumber
\\
\end{eqnarray}
Since 
\begin{eqnarray}
1 &=& \sum _x {\rm Tr}[ P_x \mathcal{U}^{\rm noisy}(\rho _{\rm ini})]
\\
&=& \alpha + {\rm Tr}[ \rho_{\rm faulty}],
\end{eqnarray}
we have $1-\alpha = {\rm Tr}[\rho_{\rm faulty}]$.
Therefore, the error is bounded by
\begin{eqnarray}
\| p(x)-p^{\rm noisy}(x) \|_{l_1} 
&\leq&  (1-\alpha) \| p(x)\|_{l_1} + \|{\rm Tr}[P_x \rho _{\rm faulty} ]\|_{l_1},
\nonumber \\
&&
\\
&\leq& 2 \| \rho _{\rm faulty} \|_{1}
\\
&\leq & 2
\prod _{k} (1-\epsilon_k)  \sum _{\{ \eta _k \} | {\rm faulty}}  \left(\frac{2\epsilon_k}{1-\epsilon_k}\right)^{\eta _k},
\nonumber \\
\label{eq:01}
\end{eqnarray}
where $\| \cdot \|_1$ indicates the operator 1-norm, 
and we used that $\| \mE_k \| _{\diamond} < 2\epsilon _k$ and $\|\mU_k \|_{\diamond}=1$.
If the system is designed fault-tolerantly, and
if the noise strength $\epsilon_k$ is smaller than 
a certain threshold value,
then the r.h.s. of Eq.~(\ref{eq:01}) is upper-bounded by 
an arbitrarily exponentially small value,
which we call the standard threshold theorem.

\section{Postselected threshold theorem}
\label{sec:02}
Next we consider the case with postselection.
In this case, we introduce another two measurement ports corresponding to
$y$ and $z$ ($\in \{ 0,1\} $).
The variable $y$ is employed as a postselection register for postBQP=PP argument~\cite{postBQP}.
The variable $z$ is used to postselect the events where
no syndrome measurement suggests an occurrence of an error.
Then we have 
\begin{eqnarray}
p(x,y,z) &=& {\rm Tr}[P_{x,y} Q_z \mU (\rho _{\rm ini}) ],
\\
p^{\rm noisy}(x,y,z) &=& {\rm Tr}[P_{x,y} Q_z \mU^{\rm nosiy} (\rho _{\rm ini}) ],
\end{eqnarray}
where $P_{x,y}$ and $Q_z$ are projectors corresponding to $(x,y)$ and $z$,
respectively.
Since, in the ideal case with $\mU$, $z$ is always zero, 
we have $ p(x,y,0) \equiv \bar p (x,y)$ and $p(x,y,1)=0$.
Now our goal here is to simulate $\bar p(x,y)$ 
by using postselected noisy quantum computation $p^{\rm noisy}(x,y|z=0)$
with an exponentially small additive error. 
Similarly to the standard threshold theorem,
we evaluate the error $\Delta$ between $\bar p(x,y)$ and $p^{\rm noisy}(x,y|z=0)$:
\begin{eqnarray}
\Delta&\equiv&\| \bar p(x,y) - p^{\rm noisy}(x,y|z=0)\|_{l_1}
\\
&=& \| \bar p(x,y) - {\rm Tr}[P_{x,y} Q_z (\rho _{\rm sparse} + \rho _{\rm faulty}) ]/q_{z=0} ] \|_{l_1},
\nonumber \\
\end{eqnarray}
where $q_{z=0} \equiv {\rm Tr}[Q_z (\rho _{\rm sparse} + \rho _{\rm faulty})]$
is the probability to postselect the null syndrome measurements.
Moreover, the sparse and faulty sets are 
redefined such that under the postselection of $z=0$
the operators in the sparse set 
results in the correct probability distribution:
\begin{eqnarray}
\bar p(x,y) &\propto& {\rm Tr}[P_{x,y} Q_z \rho _{\rm sparse}]/q_{z=0}
\\
&\equiv& \beta \bar p(x,y).
\end{eqnarray}
Since we have 
\begin{eqnarray}
1 &=& \sum _{x,y} {\rm Tr}[P_{x,z} Q_z \rho _{\rm sparse}]/q_{z=0}
\\
&=& \beta +  {\rm Tr}[ Q_z \rho _{\rm sparse} Q_z]/q_{z=0},
\end{eqnarray}
we obtain $1-\beta = {\rm Tr}[ Q_z \rho _{\rm sparse} Q_z]/q_{z=0}$.
Then we obtain a similar bound 
on the error between the ideal probability distribution 
and the postselected noisy probability distribution:
\begin{eqnarray}
\Delta&=& \| (1-\beta )\bar p(x,y) - {\rm Tr}[P_{x,y} Q_z \rho _{\rm faulty} ]/q_{z=0} ] \|_{l_1}
\\
&\leq & (1-\beta) + \| Q_z \rho _{\rm faulty} Q_z \| _{1}/q_{z=0}
\\
& \leq & 2\|  \rho _{\rm faulty} \| _{1}/q_{z=0}.
\nonumber \\
\end{eqnarray}
To proceed further calculation,
we assume that $\mE_k$ is a CPTP map,
i.e., the noise $\mN_k$ is a stochastic noise. 
In this case, we have $\| \mE_k \| = \epsilon _k$.
Moreover, since both $\rho _{\rm sparse}$ 
and $\rho _{\rm faulty}$ are density matrices, 
the postselection probability $q_{z=0}$
is lower-bounded:
\begin{eqnarray}
q_{z=0} &>& {\rm Tr}[Q_z \rho _{\rm sparse} Q_z]
\\
&>& {\rm Tr} [Q_z \prod _k(1-\epsilon_k)\mU(\rho_{\rm ini})]
\\
&=&\prod_{k} (1-\epsilon_k).
\end{eqnarray}
Thus the error $\Delta$ is again upper bounded as follows:
\begin{eqnarray}
\Delta < 2\sum _{\{ \eta _k \}| {\rm faulty}} \left(\frac{\epsilon_k}{1-\epsilon_k}\right)^{\eta _k}.
\label{eq:02}
\end{eqnarray}
If the system is designed fault-tolerantly and if 
$\epsilon_k$ is smaller than a certain constant value,
the r.h.s. of Eq.~(\ref{eq:02}) is 
upper-bounded by an exponentially small value.
Therefore we have the following postselected 
threshold theorem:
\begin{theo}[postselected threshold theorem]
\label{theorem:01}
Suppose noise is given as a stochastic one 
$\mN_k = (1-\epsilon_k) \mI + \mE_k$.
If the noise strength $\epsilon _k$ is smaller than
a certain threshold value,
we can simulate a probability distribution $\bar p(x,y)$ 
of an arbitrary universal quantum computation (uniformly generated polynomial-time 
quantum circuits) with 
an exponentially small additive error by using 
postselected noisy probability distribution 
$p(x,y|z=0)$.
\end{theo}
(The proof has been shown already in the above.)

Furthermore,
we can also show that simulation of $\bar p(x,y)$ 
with an exponentially small additive error is 
actually enough to show hardness of a sampling according to $p^{\rm noisy}(x,y,z)$ (see also Ref.~\cite{IQP3}):
\begin{lemm}
\label{lemma:01}
Let $C_\omega$ and $p_w(x,y)$ be a uniformly generated polynomial-time
quantum circuit and its probability distribution,
respectively.
If there exists a noisy quantum circuit $\mU^{\rm noisy}$ of 
the size $N={\rm poly}(n,\kappa)$
with $n$ being the size of $C_w$
such that
\begin{eqnarray}
|p_w(x,y) - p(x,y|z=0)| < e^{-\kappa},
\end{eqnarray}
then weak classical simulation
i.e., sampling according to $p(x,y,z)$
with the multiplicative error $c <\sqrt{2}$
is impossible unless the PH collapses to 
the third level.
\end{lemm}

Here weak classical simulation 
with a multiplicative error $c$
means that 
the classical sampling of $(x,y,...)$
according to the probability distribution 
$p^{\rm samp}(x,y,...)$ 
that satisfies
\begin{eqnarray}
 (1/c) p(x,y,...)< p^{\rm samp}(x,y,...) < c p(x,y,...),
\end{eqnarray}
\\
{\it Proof:}
A language $L$ is in the class
postBQP iff there 
exists a uniform family 
of postselected quantum circuits $\{C_\omega \}$
with a decision port $x$ and a postselection port $y$
such that 
\begin{eqnarray}
\textrm{if } \omega \in L, 
p_\omega (x |  y=0) \geq 1/2 +\delta 
\\
\textrm{if } \omega \notin L, 
p_\omega (x |  y=0) \leq 1/2 -\delta ,
\end{eqnarray}
where $\delta$ can be chosen arbitrary such that $0 <\delta <1/2$.
Note that 
without loss of generality
we can assume the probability to obtain $y=0$ 
is bounded, $p_{\omega}(y=0) > 2^{-6n-4}$ as shown in Ref.~\cite{IQP3}.
Now we have
\begin{eqnarray}
&&\left| 
p(x|y =0,z=0)-p_w(x|y=0)
\right|
\\
&< &
\left|
p(x,y|z=0)\left(
\frac{1}{p(y=0|z=0)}
-
\frac{1}{p_{\omega}(y=0)}
\right)
\right|
\nonumber \\
&&+
\left|
\frac{p(x,y|z=0)-p_{\omega}(x,y)}{p_{\omega}(y=0)}
\right|
\\
&<&
\frac{2e^{-\kappa}}{p(y=0|z=0)p_{\omega}(z=0)}
+
\frac{e^{-\kappa}}{p_{\omega}(z=0)}
\\
&<&
\frac{2e^{-\kappa}}{(p_{\omega}(y=0)-e^{-\kappa})p_{\omega}(y=0)}
+
\frac{e^{-\kappa}}{p_{\omega}(y=0)}.
\end{eqnarray}
Since $p_{\omega}(y=0)>2^{-6n-4}$,
we can choose $\kappa = {\rm poly}(n)$
such that  
$\left| 
p(x|y =0,z=0)-p_{\omega}(x|y_1=0)
\right| <1/2$.
The resultant size of the noisy quantum 
circuit is still polynomial in $n$.
From the definition (robustness against the bounded error) 
of the class postBQP (as same as postBQP),
the postselected noisy quantum circuit 
can decide problems in postBQP=PP
(recall that we can freely choose $0<\delta <1/2$).
Thus postselected quantum computation 
of such noisy quantum circuits is 
as hard as PP,
and hence cannot be weakly simulated with 
the multiplicative error $c <\sqrt{2}$
unless the PH collapses to the third level.
\\
\hfill $\square$

We have considered 
an approximated sampling with an multiplicative error $c$
in Lemma~\ref{lemma:01}.
Approximation 
with the constant multiplicative error 
imposes a stronger requirement on classical computers
than the constant additive error with $l_1$-norm~\cite{IQP4,IQP5}.
However, all imperfections including measurements 
are taken into $\mN_k$ on unitary gates $\mU_k$.
Since here we are interested whether or not
the outputs of the actual noisy experimental device 
possess quantum supremacy or not,
an exact sampling, i.e. $c=1$, is still enough for our purpose.
Note also that strong simulation, i.e., 
a calculation of a probability distribution $p(x)$ for a given $x$,
with a constant multiplicative error is too strong notion of
classical simulation, and hence it is much harder than
what the actual experimental device does.
However, an exact weak simulation, in which we are interested, 
is what the actual experimental device does.

By combining Theorem~\ref{theorem:01} and Lemma~\ref{lemma:01},
we obtain the following threshold theorem of quantum supremacy:
\begin{theo}[Threshold of Quantum Supremacy]
Suppose noise is given as a stochastic one $\mN _k = (1-\epsilon _k)\mI
+ \mE_k$ with $\| \mE \|_{\diamond} = \epsilon _k$.
Each unitary gate $\mU_k$ chosen from a universal 
set of gates is followed by such a noise $\mN_k$.
If the noise strength $\epsilon _k$ 
is smaller than a certain constant value,
an efficient weak classical simulation
of the sampling according to the noisy quantum circuit 
$p(x,y,z,...)={\rm Tr}[P_{x,y,z,...} \mathcal{U}^{\rm noisy}(\rho _{\rm ini})]$
is impossible unless the PH collapses to the third level.
\end{theo}

The above theorem indicates that 
even noisy sampling using noisy quantum circuits 
can have a power to exhibit quantum supremacy 
against a classical simulation of them.
One of the main benefit to 
employ the threshold theorem of quantum supremacy
is that we can show quantum supremacy experimentally
in a very noisy region where the noise strength is 
above the standard threshold, which we call a pre-threshold region.
In the following, we apply the above theorem
for two prototypical cases: concatenated fault-tolerant quantum computation~\cite{AharonovBen-OrThreshold,PreskillThreshold,KitaevThreshold,KnillThreshold} 
and topological fault-tolerant quantum computation with the surface codes~\cite{RaussendorfAnn,RaussendorfNJP,RaussendorfPRL}.

\section{Case I: concatenated quantum computation}
\label{sec:03}
To obtain a further insight,
we first consider a concatenated fault-tolerant quantum computation.
Suppose each fault-tolerant logical gate at the concatenation level $l$ consists 
of at most $M$ logical gates of the level $(l-1)$.
If we employ a quantum error correction code of a distance $d$,
any of at most $t\equiv \lfloor d-1/2\rfloor$ errors never causes a logical error.
At a concatenation level $l$,
the faulty operator is bounded by
\begin{eqnarray}
\epsilon^{(l)} &\equiv&
\sum_{r=t+1}^M
\left(\begin{array}{c}
M
\\
r
\end{array}\right) (\epsilon^{(l-1)}
)^{r} (1-\epsilon^{(l-1)})^{M-r}
\\
&\leq& C (\epsilon^{(l-1)})^{t+1}.
\end{eqnarray}
By considering the concatenation,
$\| \rho _{\rm fauly}\|_{1}$ is bounded in terms of $\epsilon \equiv \max_k \epsilon _k$ as follows:
\begin{eqnarray}
\| \rho _{\rm fauly}\|_{1} < (C^{1/t} \epsilon) ^{(t+1) ^{l}}/C^{1/t},
\end{eqnarray}
where $l$ is the number of concatenation levels
and chosen to be logarithm in the size of computation.
The threshold value is given roughly by 
$1/C^{1/t}$. 
On the other hand if we apply the threshold theorem for 
quantum supremacy,
the faulty operator is bounded at each concatenation level
as follows:
\begin{eqnarray}
\epsilon^{(l)}
=\sum_{r=d}^M
\left(\begin{array}{c}
M
\\
r
\end{array}\right) (\epsilon^{(l-1)}
)^{r} (1-\epsilon^{(l-1)})^{M-r}
\leq C' (\epsilon^{(l-1)})^{d}.
\nonumber \\
\end{eqnarray}
Similarly to the previous case,
the threshold is given roughly given by 
$1/{C'}^{1/(d-1)}$.
For example, let us take $d=3$ and $t=1$, 
and assume $M \gg 1$.
In the leading order, 
$C\sim M^2/2$ and $C' \sim M^3/6$.
The threshold of quantum supremacy
 $\epsilon _{\rm th}\sim \sqrt{6}/M^{3/2}$ is 
improved from  that $\epsilon _{\rm th}\sim 2/M^2$ for universal quantum computation 
by a factor of $O(\sqrt{M})$.

\section{Case II: fault-tolerant quantum computation
with the surface codes}
\label{sec:04}
Next we will consider topologically protected fault-tolerant quantum computation
with the surface code~\cite{RaussendorfAnn,RaussendorfNJP,RaussendorfPRL,Fowler09,FujiiText} to obtain 
a practical threshold value for quantum supremacy.
Quantum error correction using the surface code
with imperfect syndrome measurements
is governed on primal and dual cubic lattices (see Ref.~\cite{FujiiText} for a detailed review). 
Below we consider the primal cubic lattice only,
by assuming error correction is done on primal and dual cubic lattices independently,
which results in an underestimate of the threshold.
Then, errors are assigned on edges of 
the cubic lattice as an error chain, 
and the errors are detected 
at the boundary of the error chain.
If the error and recovery chains result in 
a topologically nontrivial cycle, 
the error correction fails.
In the topologically protected region, the defects representing logical qubits 
are designed such that
the nontrivial cycle consists of 
a connected chain of length longer than $d$.
Around the singular qubit for non-Clifford operations,
we have to take into account nontrivial cycles 
of length shorter than $d$ too.

Let us first consider the phenomenological noise model,
where the errors are distributed independently and identically 
on each edge with probability $\epsilon$.
Now $\rho _{\rm fail}$ is divided into 
two parts $\rho_{\rm top}$ and $\rho _{\rm sin}$,
which correspond to the errors in the topologically protected region and 
others originated from the errors around the singular qubits,
respectively.
Since we can postselect the null syndrome measurements, 
$\rho _{\rm top}$ is attributed from 
the errors on the connected chain of length longer than $d$
(the error chain of length shorter than $d$ always results in an erroneous syndrome
in the topologically protected region, and hence is postselected).
Therefore, we have 
\begin{eqnarray}
\| \rho _{\rm top} \|_1 \leq 
{\rm poly}(n)\sum _{l=d} C_l \left(\frac{\epsilon}{1-\epsilon}\right)^{l},
\label{eq:03}
\end{eqnarray}
where $n$ is the size of the quantum computation, and
$C_l < (6/5)5^l$ is the number of self-avoiding walks~\cite{Dennis} of length $l$. 
Apparently, this converges to zero if $\epsilon/(1-\epsilon) <1/5$.
Note that while the resultant threshold $\epsilon = 0.167$
is somehow underestimated by the above analytical treatment, 
it is much higher than the standard threshold $0.0293-0.033$~\cite{Dennis,Wang,Ohno} in the 
topologically protected region calculated from numerical simulations.
Therefore, the protection around singular qubit becomes important.
Around the singular qubits, the logical error is bounded by
\begin{eqnarray}
 \sum _{l=1}^{d} C'_l \left(\frac{\epsilon}{1-\epsilon}\right)^{l},
 \label{eq:04}
\end{eqnarray}
where $C'_l$ is the number of the self-avoiding walks 
that result in the logical errors of length shorter than $d$
around the singular qubits (the error of length longer than $d$ is taken in $\rho_{\rm topo}$). 
In Ref.~\cite{IQP3} (Tab. 1),
$C'_l$ is counted rigorously up to $l=14$.
This tells us that if $\epsilon/(1-\epsilon) < 0.134$ ($\epsilon <0.118$)
the amount of errors on each singular qubit 
becomes smaller than $0.146 =(1-\sqrt{2}/2)/2$, the threshold of 
the magic state distillation~\cite{MSD1,MSD2}.
Therefore $\| \rho _{\rm sim}\|_1$ converges to zero if $\epsilon <0.118$.
Accordingly, we have 
the threshold $\epsilon = 0.118$ for quantum supremacy
with the surface code under the stochastic phenomenological noise model,
which is much higher than the standard threshold $2.93\%-3.3\%$ for universal fault-tolerant quantum computation.

\begin{figure}[t]
\begin{center}
\includegraphics[width=9cm]{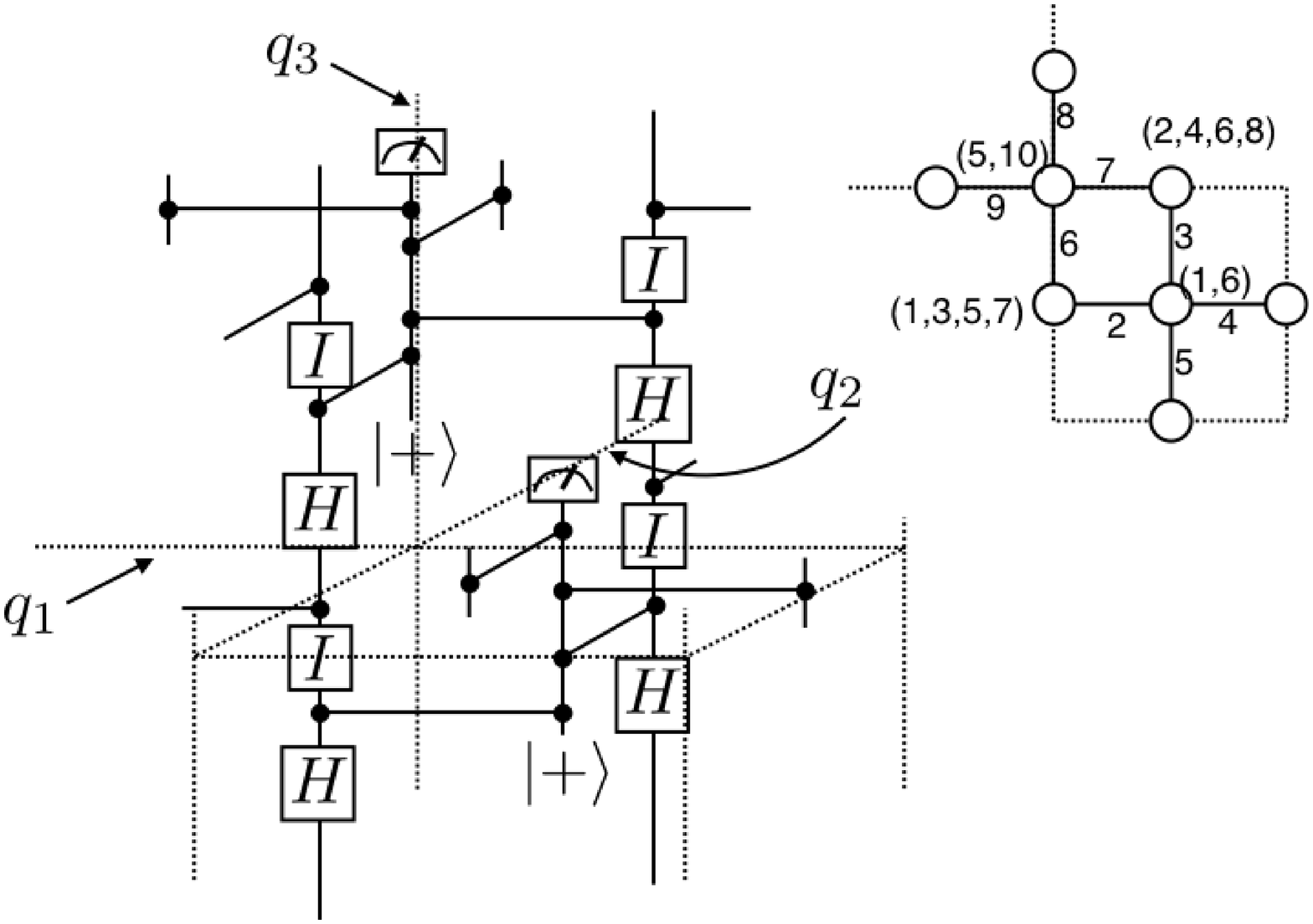}
\end{center}
%\captionsetup{justification=raggedright,singlelinecheck=false}
\caption{The depth-8 circuit for syndrome measurements of the surface code.
The top view is also shown right. An error is assigned on each edge 
with probabilities $q_1$, $q_2$, and $q_3$ independently. In addition, 
correlated errors occur with probabilities $q_{1,2}$, $q_{2,3}$, and $q_{3,1}$
on the two connected edges.}
\label{fig}
\end{figure}
\begin{figure}[t]
\begin{center}
\includegraphics[width=8cm]{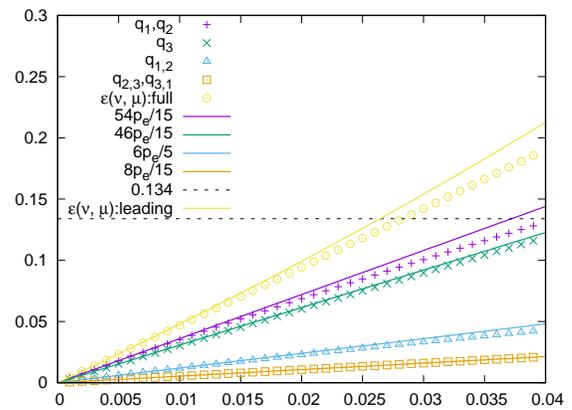}
\end{center}
%\captionsetup{justification=raggedright,singlelinecheck=false}
\caption{The parameters for the error distribution, $q_1$, $q_2$, $q_3$, 
$q_{1,2}$ , $q_{2,3}$, and $q_{3,1}$, and the effective single-qubit error probability 
$\epsilon(\nu,\mu)$ are shown in the leading order (lines) and to all orders (points)
as functions of $p_e$. 
The dashed line shows the threshold $13.4\%$ 
for the effective single-qubit error probability.}
\label{fig:02}
\end{figure}

Finally, we derive a noise threshold of quantum supremacy
in the circuit-based noise model.
Specifically, we employ a circuit shown in Fig.~\ref{fig} for the syndrome measurements 
of the surface code.
 (Contrast to the circuit in Ref.~\cite{RaussendorfPRL}, this circuit is not 
the lowest depth one. However, this setup is convenient to model the correlated errors.)
We take the standard depolarizing noise for single- and 
two-qubit gates:
\begin{eqnarray}
\mathcal{N}^{(1)} &=& [I] + \sum _{A=X,Y,Z} (p_1/3)[A] 
\\
\mathcal{N}^{(2)} &=& [I] + \sum _{A,B=X,Y,Z \backslash (A,B)=(I,I)} (p_2/15) [A\otimes B],
\end{eqnarray}
where we use the notation $[W]\rho \equiv W \rho W^{\dag}$.
The state preparations and measurements
are followed and preceded by flipping the states in their bases
with probabilities $p_p$ and $p_m$, respectively.
Then, the error distribution on the cubic lattice
is characterized by single-qubit error probabilities 
$q_1$, $q_2$, and $q_3$, and two-qubit correlated 
error probabilities $q_{1,2}$, $q_{2,3}$, and $q_{3,1}$,
where the labels $1,2$ and $3$ correspond to two space-like and one time-like axes.
In the leading order, they are given by 
\begin{eqnarray}
q_1 &= &q_2=6 \frac{4p_2}{15} + 3 \frac{2p_1}{3},
\\
q_3 &=& 4 \frac{4p_2}{15} +  p_p + p_m,
\\
q_{1,2} &=& 2\frac{4p_2}{15},
\\
q_{2,3} &=& q_{3,1} = 2 \frac{4p_2}{15}+\frac{2p_1}{3}.
\end{eqnarray}
We also numerically evaluated $q_1,q_2,q_3$ and 
$q_{1,2},q_{2,3},q_{3,1}$ to all orders
as shown in Fig.~\ref{fig:02},
which are in good agreement with 
the leading order evaluations but become
slightly smaller than them by increasing $p_e$.
%%%%
%%%%
Note that at the boundary and inside the defects
a part of the gates for the syndrome measurements are not performed,
and hence the actual error probabilities are smaller there.
Let us define $\nu = \max \{ q_1, q_2, q_3\}$ and 
$\mu = \max\{ q_{1,2}, q_{2,3}, q_{3,1}\}$.
Since we have a correlated error
on the connected two edges with probability at most $\mu$,
the probability of an error chain of the length $l$ (in Eqs.~(\ref{eq:03}) and (\ref{eq:04}))
is now replaced by 
\begin{eqnarray}
&&C_l \sum _{k=0}^{\lfloor l/2 \rfloor} 
\left(
\begin{array}{c}
	\lfloor l/2 \rfloor
\\
k
\end{array} 
\right)
2^k
\left(\frac{\nu}{1-\nu}\right)^{l-k}
 \left(\frac{\mu}{1-\mu}\right)^{k}
 \label{eq:06}
 \\
 &<&
C_l \left(\frac{\nu}{1-\nu}\right)^{l-\lfloor l/2 \rfloor}
\left[ \left(\frac{\nu}{1-\nu}\right)+
 \left(\frac{2\mu}{1-\mu}\right) \right]^{\lfloor l/2 \rfloor}.
 \label{eq:05}
\end{eqnarray}
Equation~(\ref{eq:06}) reads as follows.
The edges on the chain of the length $l$
are labeled from $1$ to $l$.
For $k=0,1,...,\lfloor l/2 \rfloor$,
$k$ correlated errors are chosen 
from the $\lfloor l/2 \rfloor$ edges labeled by odd numbers.
Then, each of the chosen edges can correlate
two neighboring  even number edges.

Equation~(\ref{eq:05}) can be viewed as a stochastic phenomenological noise model
with an effective single-qubit error probability $\epsilon(\nu,\mu)$,
\begin{eqnarray}
\epsilon (\nu,\mu)\equiv \left(\frac{\nu}{1-\nu}\right)^{1/2}
\left[ \left(\frac{\nu}{1-\nu}\right)+
 \left(\frac{2\mu}{1-\mu}\right) \right]^{1/2}.
\end{eqnarray}
Therefore, similarly to the previous argument,
if $\epsilon (\nu,\mu) < 0.134$,
$\| \rho _{\rm faulty}\|_1$ decreases exponentially.	
By setting $p_{e}=p_1=p_2=p_p=p_m$,
$\nu = 54p_e/15$ and $\mu = 6p_e/5$ in the leading order,
which results in the threshold $p_e=2.64\%$.
If employ the all-order evaluations as shown in Fig.~\ref{fig:02},
the threshold is slightly improved to $p_e=2.84\%$.
The obtained thresholds 
for quantum supremacy are again much higher than the standard threshold $0.75\%$ for 
universal fault-tolerant quantum computation.
Note that here the errors on the singular qubits 
are far overestimate. Some errors on the singular qubits
have correlation with the errors on the dual cubic lattice,
and hence can be postselected. If we consider the correlation
between the errors on the primal and dual cubic lattices,
the threshold of quantum supremacy would be further improved.

While the standard threshold $0.75\%$ is limited by 
the threshold in the topologically protected region,
the postselected threshold of quantum supremacy 
$2.84\%$ is determined purely from the limitation of 
magic state distillation.
Namely, quantum supremacy in the noisy quantum circuits 
is originated from distillability of the magic state.

\section{Conclusion and discussion}
\label{sec:05}
Here we have derived the threshold theorem of quantum supremacy
with noisy quantum circuits in the pre-threshold region.
While we employed noisy but universal set of gates here,
it would be interesting to apply the theorem 
for non-universal quantum computational models
such as BosonSampling, IQP, and DQC1 (see Ref.~\cite{IQP3} in the case of noisy commuting circuits).
In the case of BosonSampling,
if we want to show universality under postselection,
we have to take non-deterministic gates into account.
Fault-tolerant linear optical quantum computation~\cite{KLM,Nielsen04,DawsonNielsen06,Kok07,LiBenjamin,FujiiTokunaga}
would be employed even in this case.
On the other hand, in the case of DQC1~\cite{MorimaeFF,KKMNTT},
the number of measurement ports seems to be too small
to perform fault-tolerant quantum computation.
This problem might be avoided by employing polynomially many 
measurements, but still it is quite 
nontrivial to construct a fault-tolerant circuit 
by using completely randomized ancilla states and postselection.
Contrast to BosonSampling and IQP,
the output of DQC1, the normalized trace of a unitary operator,
appears ubiquitously in physics and has a lot of applications,
such as spectral density estimation~\cite{KnillLaflamme}, 
testing integrability~\cite{integrability}, calculation of fidelity decay~\cite{fidelity_decay}, and approximation of the Jones 
and HOMFLY polynomials~\cite{ShorJordan08,Passante09,JordanWocjan09}.
Fault-tolerance of quantum supremacy in DQC1 
is an important open problem.

We have considered hardness of 
an exact (or a constant multiplicative approximation) 
weak classical simulation of the noisy quantum circuits 
to know whether or not the outputs of the actual experimental 
device possess quantum supremacy.
It would be interesting to see how noise tolerance 
changes if we change the notion of approximation to 
the additive one with $l_1$-norm~\cite{IQP4,IQP5,Boixo16},
which provides an advantage to classical computers 
making their quantum targets relaxed.
Is there still surviving quantum supremacy
of pre-threshold noisy quantum circuits even in such 
a setting?

\section*{Acknowledgements}
KF is supported by KAKENHI No.16H02211 and PRESTO, JST.
This work is supported by CREST, JST and ERATO, JST.

\end{document}